\documentclass[conference]{IEEEtran}\IEEEoverridecommandlockouts

\pdfoutput=1
\usepackage{cite}
\usepackage{graphicx,times,epsfig,amssymb,rawfonts,subfigure,array,url}
\usepackage[cmex10]{amsmath}
\usepackage{balance,float}
\newtheorem{theorem}{Theorem}[section]

\floatstyle{plain}
\newfloat{longeqn}{b}{lop}

\newtheorem{remark}{Remark}[section]

\begin{document}
\title{On Achievable Rate Regions for Half-Duplex Causal Cognitive Radio Channels
  \thanks{This work was supported in part by the National
    Science Foundation under Grant CNS-0626863.}}

\author{\IEEEauthorblockN{Debdeep Chatterjee and Tan F. Wong}
\IEEEauthorblockA{Wireless Information Networking Group, Department of ECE\\
University of Florida, Gainesville, Florida 32611\\
Email: \texttt {debdeep@ufl.edu,~twong@ece.ufl.edu}}
\and
\IEEEauthorblockN{Ozgur Oyman}
\IEEEauthorblockA{Intel Labs\\
Santa Clara, CA 95054\\
Email: \texttt{ozgur.oyman@intel.com}}}

\maketitle

\newcounter{MYtempeqncnt}

\begin{abstract}
Coding for the causal cognitive radio channel, with the cognitive source subjected to a half-duplex constraint, is studied. A discrete memoryless channel model incorporating the half-duplex constraint is presented, and a new achievable rate region is derived for this channel. It is proved that this rate region contains the previously known causal achievable rate region of~\cite{Devroye06} for Gaussian channels.
\end{abstract}

\section{Introduction}\label{sec:Intro}

The cognitive radio channel~\cite{Devroye06} is the simplest kind of the overlay form of cognitive radio networks, wherein the cognitive radio simultaneously utilizes the same spectrum as the primary user-pair for its own data transmission. 
The cognitive radio channel with the cognitive source having non-causal knowledge of the primary message have been considered in many recent information theoretic works. Various achievable rate regions for the non-causal case have been proposed in~\cite{Devroye06, Maric08, Jiang08, Cao09, Rini09}, etc. 

In real deployments, some resources (in time or frequency) need to be expended by the system for the cognitive source to acquire the primary message, and this overhead should be explicitly modeled to obtain more realistic coding schemes and rate regions. In~\cite{Devroye06}, the authors consider half-duplex operation of the cognitive source, and propose four two-phase protocols. 
On the other hand, in~\cite{Seyedmehdi07}, a full-duplex operation of the cognitive source is assumed, and block Markov SPC along with sliding-window decoding, and rate-splitting for the two messages are used to obtain an achievable rate region. In~\cite{Cao09}, the causal scenario is considered from a Z interference channel (ZIC) perspective, wherein the primary destination does not experience any interference from the secondary transmission.

Recently, we derived a new achievable rate region for the full-duplex causal cognitive radio channel in~\cite{Chatterjee09}. 
In this work, we consider the causal cognitive radio channel, wherein the cognitive source is subjected to the half-duplex constraint. First, we present  a discrete memoryless channel model for the half-duplex causal cognitive radio channel (HD-CCRC), and then propose a generalized  coding scheme for this channel. 
It is also proved that the new rate region contains the previously known rate region of~\cite{Devroye06} for the Gaussian HD-CCRC.

\vspace{-.1cm}
\section{The Discrete Memoryless Channel Model}\label{sec:Channel_Model}

\begin{figure}[tb]
\begin{center}
\includegraphics[scale=0.465]{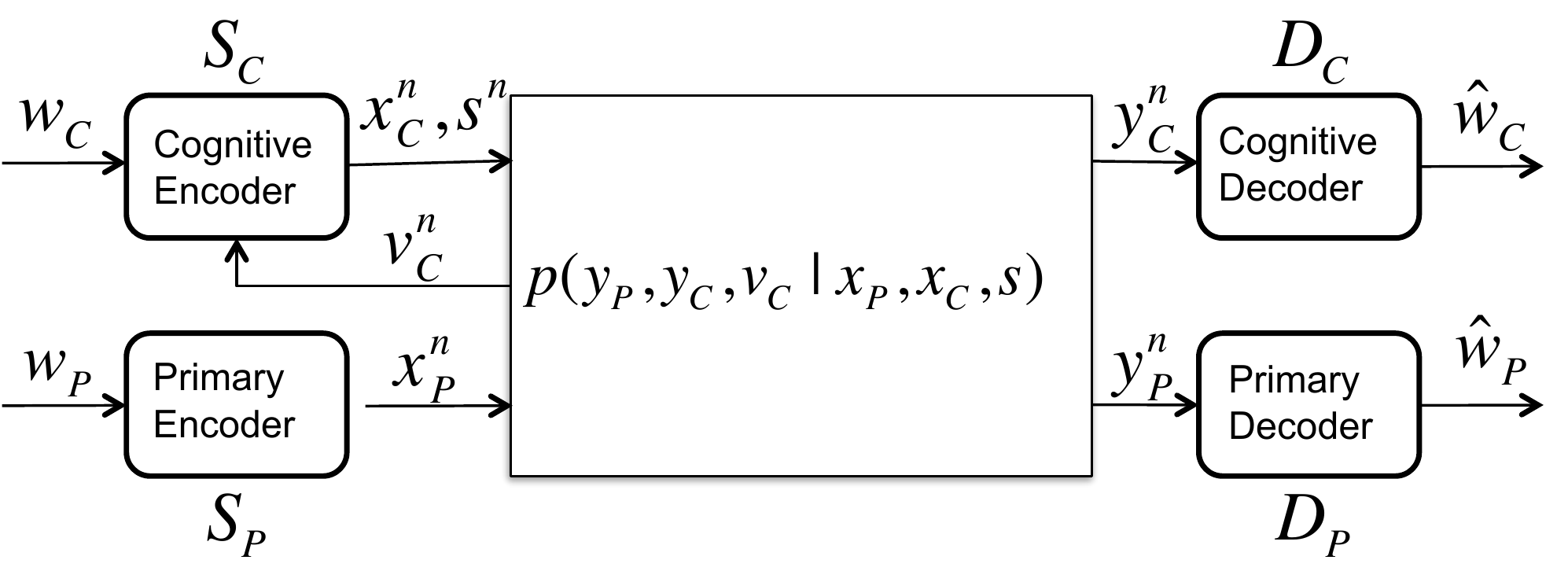}
\end{center}
\caption{The discrete memoryless HD-CCRC.}
\label{fig:dmc_HDCCRC}
\vspace{-.5cm}
\end{figure}

The HD-CCRC  is depicted in Fig.~\ref{fig:dmc_HDCCRC}, wherein the primary source node $S_P$ intends to transmit information to its destination node $D_P$.  A cognitive (or secondary) source-destination pair, $S_{C}$ and $D_C$, wishes to communicate as well, with $S_C$ having its own information to transmit to $D_{C}$. The primary message is only causally available at $S_C$. To incorporate the half-duplex constraint for the discrete memoryless channel model, we consider a second input at $S_C$, $S$, to indicate the state of $S_C$ --- listening or transmitting.

With this, the channel transition probability is determined by the state of the cognitive source as follows:
\vspace{-.25cm}
\begin{eqnarray}
&&\hspace{-1.5cm}p(y_P, y_C, v_C | x_P, x_C, s) =\nonumber\\
&&\hspace{.5cm}\begin{cases}
p(y_P, y_C, v_C | x_P) &\text{if $s=l$}\\
p(y_P, y_C | x_P, x_C)\delta_{e}(v_C) &\text{if $s=t$,}
\end{cases}\label{eqn:DMC_HD-CCR_ch_prob}
\end{eqnarray}
\noindent where $e$ denotes an erasure at $S_C$, and $\delta_e(v_C)=1$ if $v_C=e$ and $0$ otherwise. To incorporate the fact that $S_C$ cannot transmit when in the listening state, we restrict the joint probability distribution of the inputs as
$p(x_P, x_C, s) = p(x_P | s=l) \delta_{\phi}(x_C) p(s=l) + p(x_P, x_C | s=t)p(s=t)$,
where $\phi$ is the ``null'' symbol.

In $n$ channel uses, $S_P$ has message $w_P\in\{1, 2, \cdots, 2^{nR_P}\}$ to transmit to $D_P$, while $S_{C}$ has message $w_{C}\in\{1, 2, \cdots, 2^{nR_{C}}\}$ to transmit to $D_{C}$. Let $\mathcal{X}_P, \mathcal{X}_C, \mathcal{S}$, and $\mathcal{V}_C, \mathcal{Y}_P,\mathcal{Y}_C$ be the input and output alphabets respectively. Further, $\mathcal{S}=\{l, t\}$. A rate pair $(R_P,R_C)$ is achievable if there exist an encoding function for $S_P$, $X_P^n=f_P(w_P),~f_P:\{1, 2, \cdots, 2^{nR_P}\} \to \mathcal{X}_P^n$, and a sequence of encoding functions for $S_C$, $(X_C^n, S^n)=f_C^n(w_C, V_C^{n-1})$ with $(X_{Ci}, S_i)=f_{Ci}(w_C, V_C^{i-1}),~f_{Ci}:\{1, 2, \cdots, 2^{nR_C}\}\times\mathcal{V}_C^{i-1} \to \mathcal{X}_C\times \mathcal{S}$, and corresponding decoding functions $\hat{w}_P=g_P(Y_P^n),~g_P:\mathcal{Y}_P^n \to \{1, 2, \cdots, 2^{nR_P}\}$ and $\hat{w}_C=g_C(Y_C^n),~g_C:\mathcal{Y}_C^n \to \{1, 2, \cdots, 2^{nR_C}\}$ such that the average probability of error $P_e^{(n)}=\max \{P_{e,P}^{(n)}, P_{e,C}^{(n)} \} \to 0$, where $\displaystyle P_{e,M}^{(n)}=\frac{1}{2^{n(R_P+R_C)}}\sum_{(w_P,w_C)}\Pr\left[g_M(Y_M^n) \neq w_M | (w_P, w_C) \textnormal{ was sent}\right]$ for $M=P,C$.

\section{An Achievable Rate Region for the HD-CCRC} 
\label{sec:Causal}

First, we present a brief description of the coding scheme. In block $b\in\{1,\cdots,B\}$, $S_P$ splits the message $w_{P,b}$ as $w_{P,b}=(w_{P1,b}, w_{P2,b})$ with $w_{Pi,b}=(w_{Pico,b}, w_{Pipr,b})$ for $i=1,2$. Here, for any block, $w_{P1}$ is the message part that $S_C$ decodes and uses for its cognitive and cooperative actions, whereas $w_{P2}$ is the message part that $S_P$ directly transmits to $D_P$ when $S_C$ is in transmit mode. The subscripts $co$ and $pr$ indicate the common and private message parts respectively. While the common message parts are decoded by both destinations, the private message parts are decoded only by the intended destination. $w_{P1co,b}$ is further divided into two parts - $w_{s,b}$, that is forwarded by $S_C$ in the next block using the help of its random listen-transmit schedule~\cite{Kramer04},  and $w_{e,b}$, that is transmitted explicitly using a standard codebook.

Conditional rate-splitting~\cite{Cao09} and superposition coding are used for the above message splitting steps. For block $b\in\{1,\cdots,B\}$, $S_P$ transmits $w_{P1,b}$ during the $S_C$-listen states, and it superposes $w_{P2,b}$ onto $w_{P1,b-1}$ (using block Markov SPC) during the $S_C$-transmit states, with $w_{P1,b-1}$ acting as the \emph{resolution information} for $D_P$ and $D_C$ to decode $w_{P1}$ entirely or partially. In block $b$, $S_C$ decodes $w_{P1,b}$  from the received symbols during the listen-states. In block $b$, $S_C$ splits $w_{C,b}$ into two parts $w_{Cco,b}$ and $w_{Cpr,b}$, and conditioned on the codeword pair $(S, T_{P1co})$ for the resolution information for the common part of $w_{P1,b-1}$, it uses conditional GP binning~\cite{Maric08} to encode $w_{Cco,b}$ and $w_{Cpr,b}$ as $U_{Cco}$ and $U_{Cpr}$ respectively, against the resolution information for the private part of $w_{P1,b-1}$ ($T_{P1pr}$). It transmits a combination of the above codewords, along with the resolution information, during the $S_C$-transmit states.

Both $D_P$ and $D_C$ wait until the transmission in block $B$, and then use backward decoding
to jointly decode both common and private parts of its intended message and the common message part(s) from the interfering transmission. Note that $D_C$ performs backward decoding only to decode $w_{P1co,b-1}$ in order to take advantage of the block Markov SPC structure used to encode it. Table~\ref{tab:hd_rv_list} lists the random variables involved in the code construction along with their definitions.

\begin{table}[!ht]
\caption{Description of Random Variables in Theorem~\ref{thm:HD-causal_ach_rates}}
\label{tab:hd_rv_list}
\begin{tabular}[!ht]{c p{6cm}}\hline
Random Variable& Definition\\ \hline\hline
$S$& Listen-transmit state for $S_C$, used to encode $w_{s}$\\ \hline
$T_{P1co}$& Partial resolution information for common part ($w_e$) of primary message $w_{P1}$ (known to $S_C$)\\ \hline
$T_{P1pr}$& Resolution information for private part of primary message $w_{P1}$  (known to $S_C$)\\ \hline
$X_{P1co}$& New information for common part of primary message  $w_{P1}$  (decoded by $S_C$)\\ \hline
$X_{P1pr}$& New information for private part of primary message  $w_{P1}$ (decoded by $S_C$)\\ \hline
$X_{P2co}$& Common part of primary message  $w_{P2}$ (not decoded by $S_C$)\\ \hline
$X_{P2pr}$& Private part of primary message  $w_{P2}$ (not decoded by $S_C$)\\ \hline
$U_{Cco}$& Common part of secondary message (generated by conditional Gel'fand-Pinsker binning)\\ \hline
$U_{Cpr}$& Private part of secondary message (generated by conditional Gel'fand-Pinkser binning)\\ \hline
$X_P$& Transmitted codeword by $S_P$\\ \hline
$X_C$& Transmitted codeword by $S_C$\\ \hline
\end{tabular}
\end{table}

Let $\alpha=\Pr[S=l]$, and $\bar{\alpha} = 1-\alpha$. Owing to the half-duplex constraint to the channel model, we restrict the distributions for the codewords used in the codebook construction as follows:
\begin{subequations}
\begin{eqnarray}
&&\hspace{-.85cm} p(t_{P1co}|s=l) = \delta_{\phi}(t_{P1co}),\label{eqn:CW_restriction_a}\\
&&\hspace{-.85cm} p(t_{P1pr}|t_{P1co}, s=l) = \delta_{\phi}(t_{P1pr}),\label{eqn:CW_restriction_b}\\
&&\hspace{-.85cm} p(x_{P2co}|t_{P1co}, s=l) = \delta_{\phi}(x_{P2co}),\label{eqn:CW_restriction_c}\\
&&\hspace{-.85cm} p(x_{P2pr}|x_{P2co}, t_{P1pr}, t_{P1co}, s=l) = \delta_{\phi}(x_{P2pr}),\label{eqn:CW_restriction_d}\\
&&\hspace{-.85cm} p(u_{Cco}|t_{P1co}, s=l) = \delta_{\phi}(u_{Cco}),\label{eqn:CW_restriction_e}\\
&&\hspace{-.85cm} p(u_{Cpr}|u_{Cco}, t_{P1co}, s=l) = \delta_{\phi}(u_{Cpr}),\label{eqn:CW_restriction_f}\\
&&\hspace{-.85cm} p(x_{P1co}|t_{P1co}, s=t) = \delta_{\phi}(x_{P1co}),\label{eqn:CW_restriction_g}\\
&&\hspace{-.85cm} p(x_{P1pr}|x_{P1co}, t_{P1pr}, t_{P1co}, s=t) = \delta_{\phi}(x_{P1pr}). \label{eqn:CW_restriction_h}
\end{eqnarray}
\end{subequations}

\begin{theorem} \label{thm:HD-causal_ach_rates}
For the discrete memoryless HD-CCRC, all rate tuples $(R_P, R_{C})$, where $R_P=R_{P1} + R_{P2} = R_{P1co}+ R_{P1pr} + R_{P2co} + R_{P2pr}$, $R_{P1co}=R_{s}+R_{e}$, $R_{C}=R_{Cco}+R_{Cpr}$, with non-negative reals $R_s, R_e, R_{P1pr}, R_{P2co}, R_{P2pr}, R_{Cco}, R_{Cpr}$ satisfying
\begin{subequations}
\begin{eqnarray}
&&\hspace{-.75cm}R_{P1pr} \leq \alpha I\left(X_{P1pr}; V_C | X_{P1co}, S=l\right)\label{eqn:HD-causal_a}\\
&&\hspace{-.75cm}R_{P1}\leq \alpha I(X_{P1pr}; V_{C} | S=l)\label{eqn:HD-causal_b}\\
&&\hspace{-.75cm}R_{P2pr} \leq \bar{\alpha} I\left(X_{P2pr}; Y_P, U_{Cco} | X_{P2co}, T_{P1pr},\right. \nonumber\\
&&\hspace{4.75cm} \left. T_{P1co}, S=t\right) \label{eqn:HD-causal_c}\\
&&\hspace{-.75cm}R_{P2} \leq \bar{\alpha} I\left(X_{P2pr}; Y_P, U_{Cco} | T_{P1pr}, T_{P1co}, S=t \right)\label{eqn:HD-causal_d}\\
&&\hspace{-.75cm}R_{P2pr} + R_{Cco} \leq \bar{\alpha} I\left(X_{P2pr}, U_{Cco}; Y_P| X_{P2co}, T_{P1pr}, \right. \nonumber\\
&&\hspace{4.75cm}\left.T_{P1co}, S=t\right) \label{eqn:HD-causal_e}\\
&&\hspace{-.75cm}R_{P2} + R_{Cco} \leq \bar{\alpha} I\left(X_{P2pr}, U_{Cco}; Y_P | T_{P1pr}, T_{P1co}, S=t\right)\label{eqn:HD-causal_f}\\
&&\hspace{-.75cm}R_{P1pr} + R_{P2pr} \leq \alpha I\left(X_{P1pr}; Y_P | X_{P1co}, S=l\right) \nonumber\\
&&\hspace{.15cm} + \bar{\alpha} I\left(T_{P1pr}, X_{P2pr}; Y_P, U_{Cco} | X_{P2co}, T_{P1co}, S=t\right) \label{eqn:HD-causal_g} \\
&&\hspace{-.75cm}R_{P1pr} + R_{P2} \leq \alpha I\left(X_{P1pr}; Y_P | X_{P1co}, S=l\right) \nonumber\\
&&\hspace{.15cm} + \bar{\alpha} I\left(T_{P1pr}, X_{P2pr}; Y_P, U_{Cco} | T_{P1co}, S=t\right)\label{eqn:HD-causal_h} \\
&&\hspace{-.75cm}R_{P1pr} + R_{P2pr} + R_{Cco} \leq \alpha I\left(X_{P1pr}; Y_P | X_{P1co}, S=l\right) \nonumber\\
&&\hspace{.15cm} + \bar{\alpha} I\left(T_{P1pr}, X_{P2pr}, U_{Cco}; Y_P | X_{P2co}, T_{P1co}, S=t\right) \label{eqn:HD-causal_i}\\
&&\hspace{-.75cm}R_{P1pr} + R_{P2} + R_{Cco} \leq \alpha I\left(X_{P1pr}; Y_P | X_{P1co}, S=l\right) \nonumber\\
&&\hspace{.75cm} + \bar{\alpha} I\left(T_{P1pr}, X_{P2pr}, U_{Cco}; Y_P | T_{P1co}, S=t\right) \label{eqn:HD-causal_j}\\
&&\hspace{-.75cm}R_e + R_{P1pr} + R_{P2} + R_{Cco} \leq \alpha I\left(X_{P1pr}; Y_P | S=l\right) \nonumber\\
&&\hspace{.75cm} + \bar{\alpha} I\left(T_{P1co}, T_{P1pr}, X_{P2pr}, U_{Cco}; Y_P | S=t\right) \label{eqn:HD-causal_k}\\
&&\hspace{-.75cm}R_{P} + R_{Cco} \leq I\left(S; Y_P\right) + \alpha I\left(X_{P1pr}; Y_P | S=l\right)\nonumber\\
&&\hspace{.75cm} + \bar{\alpha} I\left(T_{P1co}, T_{P1pr}, X_{P2pr}, U_{Cco}; Y_P | S=t\right) \label{eqn:HD-causal_l}\\
&&\hspace{-.75cm}R_{Cpr} \leq \bar{\alpha} \left[ I\left(U_{Cpr}; Y_C, U_{Cco} | X_{P2co}, T_{P1co}, S=t \right) \right. \nonumber\\
&&\hspace{1.5cm}\left.- I \left( U_{Cpr}; T_{P1pr}, U_{Cco} | T_{P1co}, S=t\right) \right] \label{eqn:HD-causal_m}\\
&&\hspace{-.75cm}R_{C} \leq \bar{\alpha} \left[ I\left(U_{Cco}, U_{Cpr}; Y_C | X_{P2co}, T_{P1co}, S=t \right) \right. \nonumber\\
&&\hspace{1.5cm}\left. - I \left(U_{Cco}, U_{Cpr}; T_{P1pr} | T_{P1co}, S=t\right) \right] \label{eqn:HD-causal_n}\\
&&\hspace{-.75cm}R_{P2co} + R_{Cpr} \leq \bar{\alpha} \left[ I\left(X_{P2co}, U_{Cpr}; Y_C, U_{Cco} | T_{P1co}, S=t \right) \right. \nonumber\\
&&\hspace{1.5cm}\left. - I \left(U_{Cpr}; T_{P1pr}, U_{Cco} | T_{P1co}, S=t\right) \right] \label{eqn:HD-causal_o}\\
&&\hspace{-.75cm}R_{P2co} + R_{C} \leq \bar{\alpha} \left[ I\left(X_{P2co}, U_{Cco}, U_{Cpr}; Y_C | T_{P1co}, \right. \right. \nonumber\\
&&\hspace{.5cm}\left.\left. S=t \right) - I \left(U_{Cco}, U_{Cpr}; T_{P1pr} | T_{P1co}, S=t\right) \right] \label{eqn:HD-causal_p}\\
&&\hspace{-.75cm}R_e + R_{P2co} + R_{C} \leq \alpha I \left( X_{P1co}; Y_C | S=l \right) \nonumber\\
&&\hspace{.75cm} + \bar{\alpha} \left[ I\left(T_{P1co}, X_{P2co}, U_{Cco}, U_{Cpr}; Y_C | S=t \right) \right. \nonumber\\
&&\hspace{1.75cm}\left. - I \left( U_{Cco}, U_{Cpr}; T_{P1pr} | T_{P1co}, S=t\right) \right] \label{eqn:HD-causal_q}\\
&&\hspace{-.75cm}R_{P1co} + R_{P2co} + R_{C} \leq I\left(S;Y_C\right) + \alpha I \left( X_{P1co}; Y_C | S=l \right) \nonumber\\
&&\hspace{.75cm} + \bar{\alpha} \left[ I\left(T_{P1co}, X_{P2co}, U_{Cco}, U_{Cpr}; Y_C | S=t \right) \right. \nonumber\\
&&\hspace{1.75cm}\left. - I \left( U_{Cco}, U_{Cpr}; T_{P1pr} | T_{P1co}, S=t\right) \right] \label{eqn:HD-causal_r}
\end{eqnarray}
\end{subequations}
\noindent are achievable for some joint distribution that factors as
\begin{eqnarray*}
&&\hspace{-.75cm}p(s) p(t_{P1co}|s) p(t_{P1pr} | t_{P1co}, s) p(x_{P1co} | t_{P1co}, s) \\
&&\hspace{-.75cm}\times p\left(x_{P1pr} | x_{P1co}, t_{P1pr}, t_{P1co}, s\right) p(x_{P2co} | t_{P1co}, s) \\
&&\hspace{-.75cm}\times p\left(x_{P2pr} | x_{P2co}, t_{P1pr}, t_{P1co}, s\right) p\left(x_P | x_{P2pr}, x_{P2co}, \right. \\
&&\hspace{-.75cm}\left. x_{P1pr}, x_{P1co}, t_{P1pr}, t_{P1co}, s\right) p\left(u_{Cco} | t_{P1co}, s\right) \\
&&\hspace{-.75cm}\times p\left(u_{Cpr} | u_{Cco}, t_{P1co}, s\right)p\left(x_C | u_{Cpr}, u_{Cco}, t_{P1pr}, t_{P1co}, s\right) \\
&&\hspace{-.75cm} \times p\left(v_{C}| x_P, x_{C}, s\right) p\left(y_{P}| x_P, x_{C}, s\right) p\left(y_{C}| x_P, x_{C}, s\right), \label{eqn:joint_dist}
\end{eqnarray*}
and satisfies~\eqref{eqn:CW_restriction_a}-\eqref{eqn:CW_restriction_h}, and for which the right-hand sides of~\eqref{eqn:HD-causal_a}-\eqref{eqn:HD-causal_r} are non-negative. 
\end{theorem}

\begin{IEEEproof}\label{proof:causal_ach_rates}
Let $\mathcal{A}_{\epsilon}^n(X,Y)$ denote set of jointly $\epsilon$-typical sequences according to the distribution of random variables $X,~Y$ as induced by the same distribution used to generate the codebooks. For the sake of space, the dependence on the random variables will not be stated explicitly, and should be clear from the context.  \\
\noindent\textbf{Codebook generation:} Split the primary and cognitive users' rates as $R_{P} = R_s + R_e + R_{P1pr} + R_{P2co} + R_{P2pr}$, and $R_{C}=R_{Cco} + R_{Cpr}$ respectively. Fix a distribution $p\left(s, t_{P1co}, \right. \\
\left. t_{P1pr}, x_{P1co}, x_{P1pr}, x_{P2co}, x_{P2pr}, x_P, u_{Cco}, u_{Cpr}, x_{C}\right)$ as in Theorem~\ref{thm:HD-causal_ach_rates}.

\begin{itemize}
\item Generate $2^{nR_{s}}$ i.i.d. codewords $s^n(w'_{s})\in\mathcal{S}^n$, $w'_{s}\in\{1,\cdots,2^{nR_{s}}\}$, according to $\prod_{i=1}^n p(s_{i})$.

\item For each codeword $s^n(w'_{s})$, generate $2^{nR_{e}}$ conditionally i.i.d. codewords $t_{P1co}^n(w'_{s}, w'_{e})$, $w'_{e}\in\{1,\cdots,2^{nR_{e}}\}$, according to $\prod_{i=1}^np(t_{P1coi}|s_{i})$.


\item For each codeword pair $(s^n(w'_{s}), t_{P1co}^n(w'_s, w'_e))$, generate $2^{nR_{P1pr}}$ conditionally i.i.d. codewords $t_{P1pr}^n(w'_{s}, w'_{e}, w'_{P1pr})$, $w'_{P1pr}\in\{1,\cdots,2^{nR_{P1pr}}\}$, according to $\prod_{i=1}^np(t_{P1pri}|s_{i}, t_{P1coi})$.

\item For each codeword pair $(s^n(w'_{s}), t_{P1co}^n(w'_s, w'_e))$, generate $2^{nR_{P1co}}$ conditionally i.i.d. codewords $x_{P1co}^n(w'_{s}, w'_{e}, w_{P1co})$, $w_{P1co}\in\{1,\cdots,2^{nR_{P1co}}\}$, according to $\prod_{i=1}^np(x_{P1coi}|s_{i}, t_{P1coi})$.

\item For each codeword tuple $\left(s^n(w'_{s}), t_{P1co}^n(w'_s, w'_e),  \right. \\
\left. x^n_{P1co}(w'_s, w'_e, w_{P1co}), t_{P1pr}^n(w'_{s}, w'_{e}, w'_{P1pr})\right)$, generate $2^{nR_{P1pr}}$ conditionally i.i.d. codewords $x_{P1pr}^n\left(w'_{s}, w'_{e}, \right. \\
\left. w_{P1co}, w'_{P1pr}, w_{P1pr}\right)$, $w_{P1pr}\in \left\{1,\cdots, 2^{nR_{P1pr}}\right\}$, according to $\prod_{i=1}^np(x_{P1pri}|s_{i}, t_{P1coi}, x_{P1coi}, t_{P1pri})$.

\item For each codeword pair $(s^n(w'_{s}), t_{P1co}^n(w'_s, w'_e))$, generate $2^{nR_{P2co}}$ conditionally i.i.d. codewords $x_{P2co}^n(w'_{s}, w'_{e}, w_{P2co})$, $w_{P2co}\in\{1,\cdots,2^{nR_{P2co}}\}$, according to $\prod_{i=1}^np(x_{P2coi}|s_{i}, t_{P1coi})$.

\item For each codeword tuple $\left(s^n(w'_{s}), t_{P1co}^n(w'_s, w'_e), \right. \\
\left. x_{P2co}^n(w'_{s}, w'_{e}, w_{P2co}), t_{P1pr}^n(w'_{s}, w'_{e}, w'_{P1pr})\right)$, generate $2^{nR_{P2pr}}$ conditionally i.i.d. codewords $x_{P2pr}^n\left(w'_{s}, w'_{e}, \right. \\
\left. w_{P2co}, w'_{P1pr}, w_{P2pr}\right)$, $w_{P2pr}\in\left\{1,\cdots, 2^{nR_{P2pr}}\right\}$, according to $\prod_{i=1}^np(x_{P2pri}|s_{i}, t_{P1coi}, x_{P2coi}, t_{P1pri})$.

\item For each codeword pair $(s^n(w'_s), t_{P1co}^n(w'_s, w'_e))$, generate $2^{n(R_{Cco}+R'_{Cco})}$ i.i.d. codewords $u_{Cco}^n(w'_s, w'_e, w_{Cco}, b_{Cco})$, $w_{Cco}\in\{1,\cdots,2^{nR_{Cco}}\}$ and $b_{Cco}\in\{1,\cdots,2^{nR'_{Cco}}\}$, according to $\prod_{i=1}^n p(u_{Ccoi}| s_i, t_{P1coi})$.

\item For each codeword tuple $\left(s^n(w'_s), t_{P1co}^n(w'_s, w'_e), u_{Cco}^n\left(w'_s, w'_{e}, 
w_{Cco}, b_{Cco}\right)\right)$, generate $2^{n(R_{Cpr}+R'_{Cpr})}$ i.i.d. codewords $u_{Cpr}^n(w'_s, w'_e, w_{Cco}, b_{Cco}, w_{Cpr}, b_{Cpr})$, $w_{Cpr}\in\{1,\cdots,2^{nR_{Cpr}}\}$ and $b_{Cpr}\in\{1,\cdots,2^{nR'_{Cpr}}\}$, according to $\prod_{i=1}^n p(u_{Cpri}| s_i, t_{P1coi}, u_{Ccoi})$.

\item Generate $x_{P}^n(w'_s, w'_e, w'_{P1pr}, w_{P1co}, w_{P1pr}, w_{P2co}, w_{P2pr})$ where $x_{P}$ is a deterministic function of $s, t_{P1co}, t_{P1pr}, x_{P1co}, x_{P1pr}, x_{P2co}, x_{P2pr}$.

\item Generate $x_{C}^n(w'_s, w'_e, w'_{P1pr}, w_{Cco}, b_{Cco}, w_{Cpr}, b_{Cpr})$ where $x_{C}$ is a deterministic function of $s, t_{P1co}, t_{P1pr}, u_{Cco}, u_{Cpr}$ such that $x_{C}=\phi$ if $s=l$.

\end{itemize}

\textbf{Encoding:} At $S_P$: In block $b\in\{2,\cdots,B-1\}$, $S_P$ transmits $x_{P}^n\left(w_{s,b-1}, w_{e,b-1}, w_{P1pr,b-1}, w_{P1co,b}, w_{P1pr,b}, \right.\\
\left. w_{P2co,b}, w_{P2pr,b}\right)$. In the first block, $S_P$ transmits $x_{P}^n(1, 1, 1, w_{P1co,1}, w_{P1pr,1}, w_{P2co,1}, w_{P2pr,1})$, while in block $B$, it transmits $x_{P}^n\left(w_{s,B-1}, w_{e,B-1}, w_{P1pr,B-1}, 1, 1,\right. \\
\left. w_{P2co,B}, w_{P2pr,B}\right)$. Note that the actual rate for the primary message is $\frac{B-1}{B}(R_s+R_e+R_{P1pr}) + R_{P2co} + R_{P2pr}$, but it converges to $R_P$ as the number of blocks $B$ goes to infinity.

At $S_{C}$: In block $b\in\{1,\cdots, B\}$, to transmit $w_{Cco,b}$, $S_{C}$ searches for bin index $b_{Cco,b}$ such that 
\begin{eqnarray}
&&\hspace{-1.25cm}\left(s^n(\hat{\bar{w}}_s), t_{P1co}^n(\hat{\bar{w}}_s, \hat{\bar{w}}_e), u_{Cco}^n(\hat{\bar{w}}_s, \hat{\bar{w}}_e, w_{Cco,b}, b_{Cco,b}), \right. \nonumber\\
&&\hspace{2.25cm}\left. t_{P1pr}^n(\hat{\bar{w}}_s, \hat{\bar{w}}_e,\hat{\bar{w}}_{P1pr})\right)\in\mathcal{A}_{\epsilon}^n,\label{eqn:S_C_encoding_co}
\end{eqnarray}
where $\hat{\bar{w}}_s, \hat{\bar{w}}_e$ and $\hat{\bar{w}}_{P1pr}$ are $S_{C}$'s estimates of $w_{s,b-1}, w_{e,b-1}$ and $w_{P1pr,b-1}$ respectively from the previous block. Once $b_{Cco,b}$ is determined, it searches for a bin index $b_{Cpr, b}$ in order to transmit $w_{Cpr,b}$ such that 
\begin{eqnarray}
&&\hspace{-.5cm}\left(s^n(\hat{\bar{w}}_s), t_{P1co}^n(\hat{\bar{w}}_s, \hat{\bar{w}}_e), u_{Cco}^n(\hat{\bar{w}}_s, \hat{\bar{w}}_e, w_{Cco,b}, b_{Cco,b}), \right.\nonumber\\
&&\hspace{-.5cm} \left. u_{Cpr}^n\left(\hat{\bar{w}}_s, \hat{\bar{w}}_e, w_{Cco,b}, b_{Cco,b},w_{Cpr,b}, b_{Cpr,b}\right), \right. \nonumber\\
&&\hspace{-.5cm}\left. t_{P1pr}^n(\hat{\bar{w}}_s, \hat{\bar{w}}_e,\hat{\bar{w}}_{P1pr})\right)\in\mathcal{A}_{\epsilon}^n. \label{eqn:S_C_encoding_pr} 
\end{eqnarray}
It sets $b_{Cco,b}=1$ or $b_{Cpr,b}=1$ if the respective bin index is not found. It can be shown using arguments similar to those in~\cite{Maric08} that the probabilities of the events of $S_C$ not able to find a unique $b_{Cco,b}$ or $b_{Cpr,b}$ satisfying~\eqref{eqn:S_C_encoding_co} and~\eqref{eqn:S_C_encoding_pr} can be made arbitrarily small if the following hold true:
\begin{eqnarray*}
&&R'_{Cco} > \bar{\alpha} I(U_{Cco}; T_{P1pr} | T_{P1co}, S=t)+\epsilon_0,\\
&&R'_{Cpr} > \bar{\alpha} I(U_{Cpr}; T_{P1pr} | U_{Cco}, T_{P1co}, S=t)+\epsilon_0,
\end{eqnarray*}   
\noindent where $\epsilon_0>0$ may be arbitrarily small. $S_{C}$ transmits $x_{C}^n(w_{s,b-1}, w_{e,b-1}, w_{P1pr,b-1}, w_{Cco,b}, b_{Cco,b}, w_{Cpr,b}, b_{Cpr,b})$.

\textbf{Decoding:} At $S_{C}$: Assume that decoding till block $b-1$ has been successful. Then, in block $b$, $S_C$ knows $w_{P1co,b-1}=(w_{s,b-1}, w_{e,b-1})$ and $w_{P1pr,b-1}$. It declares that the pair $(w_{P1co,b}, w_{P1pr,b})=(\hat{\bar{w}}_{P1co},\hat{\bar{w}}_{P1pr})$ was transmitted in block $b$ if there exists a unique pair $(\hat{\bar{w}}_{P1co}, \hat{\bar{w}}_{P1pr})$
such that
\begin{eqnarray*}
&&\hspace{-.75cm}\left(s^n(w_{s,b-1}), t_{P1co}^n(w_{s,b-1}, w_{e,b-1}), t_{P1pr}^n\left(w_{s,b-1}, w_{e,b-1}, \right. \right. \\
&&\hspace{-.75cm}\left. \left. w_{P1pr,b-1}\right), x_{P1co}^n\left(w_{s,b-1}, w_{e,b-1}, \hat{\bar{w}}_{P1co}\right), x_{P1pr}^n\left(w_{s,b-1}, \right. \right. \\
&&\hspace{-.75cm} \left. \left. w_{e,b-1}, \hat{\bar{w}}_{P1co}, w_{P1pr,b-1}, \hat{\bar{w}}_{P1pr}\right), v_{C,b}^n\right) \in\mathcal{A}_{\epsilon}^n. 
\end{eqnarray*}
\noindent Else, an error is declared. It can be shown that the probability of error for this decoding step can be made arbitrarily low if~\eqref{eqn:HD-causal_a} and~\eqref{eqn:HD-causal_b} are satisfied.

At $D_P$: The primary destination $D_P$ waits until block $B$, and then performs backward decoding. We consider the decoding process using the output in block $b\in\{B-1,\cdots, 2\}$. The decoding for the first and last blocks can be seen as special cases of the above. Thus, for block $b\in\{B-1, \cdots, 2\}$, assuming that the decoding for the pair $(w_{P1co,b}, w_{P1pr,b})$ has been successful from block $b+1$, $D_P$ searches for a unique tuple $(\hat{w}_{s}, \hat{w}_{e}, \hat{w}_{P1pr}, \hat{w}_{P2co}, \hat{w}_{P2pr})$ and some tuple $(\hat{\bar{w}}_{Cco}, \hat{\bar{b}}_{Cco})$ such that
\begin{eqnarray*}
&&\hspace{-.75cm}\left(s^n(\hat{w}_{s}), t_{P1co}^n(\hat{w}_{s}, \hat{w}_{e}), t_{P1pr}^n(\hat{w}_{s}, \hat{w}_{e}, \hat{w}_{P1pr}), x_{P1co}^n\left(\hat{w}_{s}, \hat{w}_{e}, \right. \right. \\
&&\hspace{-.75cm} \left. \left. w_{P1co,b}\right), x_{P1pr}^n\left(\hat{w}_{s}, \hat{w}_{e},  w_{P1co,b},  \hat{w}_{P1pr}, w_{P1pr,b}\right), x_{P2co}^n\left(\hat{w}_{s}, \right. \right. \\
&&\hspace{-.75cm} \left. \left. \hat{w}_{e}, \hat{w}_{P2co}\right), x_{P2pr}^n\left(\hat{w}_{s}, \hat{w}_{e},  \hat{w}_{P2co},  \hat{w}_{P1pr}, \hat{w}_{P2pr}\right),
u_{Cco}^n\left(\hat{w}_{s}, \right. \right. \\
&&\hspace{-.75cm} \left. \left. \hat{w}_{e}, \hat{\bar{w}}_{Cco}, \hat{\bar{b}}_{Cco}\right), y_{P,b}^n\right)\in\mathcal{A}_{\epsilon}^n.
\end{eqnarray*}
\noindent  The error analysis for this decoding step (omitted due to space constraints) can be used to prove that, for $n$ large enough, $(\hat{w}_{s}, \hat{w}_{e}, \hat{w}_{P1pr}, \hat{w}_{P2co}, \hat{w}_{P2pr}) = (w_{s,b-1}, w_{e,b-1}, w_{P1pr,b-1}, w_{P2co,b}, w_{P2pr,b})$ with arbitrarily small probability of error if~\eqref{eqn:HD-causal_c}-\eqref{eqn:HD-causal_l} are satisfied.

At $D_{C}$: The cognitive destination $D_{C}$ also waits until block $B$, and then performs backward decoding to jointly decode the messages intended for it and the common part of the primary message. For block $b\in\{B-1, \cdots, 2\}$, $D_C$ is assumed to have successfully decoded $w_{P1co,b}$ from block $b+1$. With this knowledge, it searches for a unique tuple $(\hat{\hat{w}}_{s}, \hat{\hat{w}}_{e}, \hat{w}_{Cco}, \hat{b}_{Cco}, \hat{w}_{Cpr}, \hat{b}_{Cpr})$ and some $\hat{\hat{w}}_{P2co}$ such that 
\begin{eqnarray*}
&&\hspace{-.75cm}\left(s^n(\hat{\hat{w}}_{s}), t_{P1co}^n(\hat{\hat{w}}_{s}, \hat{\hat{w}}_{e}), x_{P1co}^n(\hat{\hat{w}}_{s}, \hat{\hat{w}}_{e}, w_{P1co,b}), \right. \\
&&\hspace{-.75cm} \left. x_{P2co}^n(\hat{\hat{w}}_{s}, \hat{\hat{w}}_{e}, \hat{\hat{w}}_{P2co}), u_{Cco}^n(\hat{\hat{w}}_{s}, \hat{\hat{w}}_{e}, \hat{w}_{Cco}, \hat{b}_{Cco}), \right. \\
&&\hspace{-.75cm}\left. u_{Cpr}^n(\hat{\hat{w}}_{s}, \hat{\hat{w}}_{e}, \hat{w}_{Cco}, \hat{b}_{Cco}, \hat{w}_{Cpr}, \hat{b}_{Cpr}), y_{C,b}^n\right)\in\mathcal{A}_{\epsilon}^n.
\end{eqnarray*}
\noindent Again, using the properties of joint typicality, it can be established that, for $n$ large enough, $(\hat{\hat{w}}_{s}, \hat{\hat{w}}_{e}, \hat{w}_{Cco}, \hat{b}_{Cco}, \hat{w}_{Cpr}, \hat{b}_{Cpr}) = \left(w_{s,b-1}, w_{e,b-1}, w_{Cco,b}, \right.\\
\left. b_{Cco,b}, w_{Cpr,b}, b_{Cpr,b}\right)$ with an arbitrarily low probability of error if~\eqref{eqn:HD-causal_m}-\eqref{eqn:HD-causal_r} are satisfied.

Thus, the constraints on the rates as given in~\eqref{eqn:HD-causal_a} - \eqref{eqn:HD-causal_r} ensure that the average probability of error at the two destinations can be driven to zero and thus, they describe an achievable rate region for the causal cognitive radio channel.
\end{IEEEproof}

\begin{remark}
\label{rem:w_P2transmission}
According to the above coding scheme, a part of the primary message ($w_{P2}$) is not decoded by $S_C$. This is different from the non-causal case. As $S_C$ cannot receive while it transmits, $S_P$ may improve its rates by transmitting ``fresh'' information directly to the destination during $S_C$-transmit states, thereby increasing the achievable rate region. 
Note that the maximum increase in the achievable rates in using a random listen-transmit schedule for $S_C$ is $1$bit. 
\end{remark}

\begin{remark}
\label{rem:convexity}
The achievable rate region described in Theorem~\ref{thm:HD-causal_ach_rates} is convex and hence, no time-sharing is required to enlarge the rate region. This can be easily proved using the Markov chain structure of the code as was used in~\cite[Lemma 5]{Csiszar78}, with the random variable $S$ in Theorem~\ref{thm:HD-causal_ach_rates} playing a role similar to that of $U$ in~\cite{Csiszar78}.
\end{remark}

\begin{remark}
\label{rem:Gaussian}
For the Gaussian channel model with a fixed listen-transmit schedule, the coding scheme of Theorem~\ref{thm:HD-causal_ach_rates} yields the same rate region as with a time-division strategy with the use of Gaussian parallel channels~\cite{Host-Madsen05}, instead of a block Markov structure, for the decoding of $w_{P1}=(w_{P1co}, w_{P1pr})$ at $D_P$ and $w_{P1co}$ at $D_C$. According to this strategy, $S_P$ transmits  $w_{P1}$ during the first time-slot while $S_C$ is in listening mode. In the second time-slot, both $S_P$ and $S_C$ encode and transmit $w_{P1}$ as a non-causal cognitive radio channel, and $S_P$ also superposes $w_{P2}$ on top of $w_{P1}$.
 Both destinations decode only at the end of the second time-slot and exploit the parallel Gaussian channel structure to decode $w_{P1}$.
\end{remark}


\section{Inclusion Of Causal Achievable Region of~\cite{Devroye06}}
\label{sec:Comp_to_Devroye06}

In~\cite{Devroye06}, an achievable rate region for the Gaussian HD-CCRC was presented. The authors proposed four protocols and the overall achievable rate region ($\mathcal{R}_0$) is given by the convex hull of the four rate regions~\cite[Theorem 5]{Devroye06}. In this section, we show that the rate region of Theorem~\ref{thm:HD-causal_ach_rates}, $\mathcal{R}$, contains $\mathcal{R}_0$. We show that an outer bound (not necessarily achievable) to the rate region presented in~\cite{Devroye06} is contained in a subspace of the achievable rate region of Theorem~\ref{thm:HD-causal_ach_rates}. 

For the non-causal cognitive radio channel (NC-CRC), the containment of the region of~\cite[Corollary 2]{Devroye06}, $\mathcal{R}_{DMT}$, in the region $\mathcal{R}_{D}$ of~\cite[Theorem 1]{Devroye07} is clear. It is shown in~\cite{Rini09} that $\mathcal{R}_{RTD}$~\cite[Theorem 1]{Rini09} contains $\mathcal{R}_{D}$. More specifically,~\cite{Rini09} shows that $\mathcal{R}_D \subseteq \mathcal{R}^{out}_{D} \subseteq \mathcal{R}^{in}_{RTD} \subseteq \mathcal{R}_{RTD}$, where $\mathcal{R}^{out}_{D}$ is obtained from $\mathcal{R}_{D}$ by removing certain rate constraints, and $\mathcal{R}^{in}_{RTD}$ is obtained from $\mathcal{R}_{RTD}$ by restricting the input distribution to match that for $\mathcal{R}_D$. The coding scheme of Theorem~\ref{thm:HD-causal_ach_rates} may be specialized to yield a rate region for the NC-CRC. Towards this, we set $S=t~\text{w.p.}~1, X_{P2co}=X_{P2pr}=\phi$, and assume that a genie provides $S_C$ with $w_{P}$. This gives us an achievable rate region $\mathcal{R}_{NC}$ for the NC-CRC. Moreover, by restricting the input distribution to independent rate-splitting and independent binning of the secondary messages (as in~\cite{Devroye06, Devroye07}) instead of conditional rate-splitting and  conditional binning at $S_C$, it can be shown using an appropriate mapping of the codebook random variables (omitted due to lack of space), that the resulting region $\mathcal{R}^{in}_{NC}$ is identical to $\mathcal{R}^{in}_{RTD}$, and hence, $\mathcal{R}_{DMT} \subseteq \mathcal{R}_{D} \subseteq \mathcal{R}^{in}_{NC}$. 

\balance

Next, we show that the rate regions obtained via each of the protocols proposed in~\cite{Devroye06} are contained in $\mathcal{R}$. Note that for all these protocols, $w_P=w_{P1},~w_{P}=(w_{Pco}, w_{Ppr})$, with rates $R_P=R_{Pco}+R_{Ppr}$, etc.  According to Protocol 1, for any choice of $\alpha$, the rate pair $(R_P, R_C)$ is achievable if
\begin{eqnarray}
&&\hspace{-.75cm} R_P \leq \frac{\alpha}{2}\left[ \log\left(1 + g_{PC}\bar{\eta}P_P\right) + \log\left(1 + \frac{\eta P_P}{1+\bar{\eta} P_P}\right)\right],  \label{eqn:Protocol1_bd1}\\
&&\hspace{-.75cm} (R^0_P, R^0_C) \in \mathcal{R}_{DMT}, ~R_C = \bar{\alpha} R^0_C,~R_{Pco} = \bar{\alpha} R^0_{Pco}, \label{eqn:Protocol1_bd2}\\
&&\hspace{-.75cm} R_{Ppr} \leq \frac{\alpha}{2} \log\left(1 + \frac{\eta P_P}{1+\bar{\eta}P_P}\right) + \bar{\alpha}R^0_{Ppr}, \label{eqn:Protocol1_bd3}
\end{eqnarray}
\noindent where $P_P$ and $P_C$ are the respective power constraints for $S_P$ and $S_C$, $g_{PC}$ is the channel gain for the $S_P\to S_C$ link, and $\eta\in[0,1]$  is the power fraction allocated for transmitting a part (same as $\bar{\alpha}$ in~\cite{Devroye06}) of $w_{Ppr}$. The direct links are assumed to have unit channel gains, $h_{PC}$ is the channel gain for the $S_P\to D_C$ link, $h_{CP}$ is that for the $S_C\to D_P$ link, and $S_C$, $D_P$, and $D_C$ experience i.i.d. AWGN of unit-variance.

Consider the region corresponding to the fixed listen-transmit schedule and using parallel Gaussian channels as in Remark~\ref{rem:Gaussian}. For the first time-slot, set the input distribution at $S_P$ as $p(x_{P1co}|s=l)p(x_{P1pr}|x_{P1co}, s=l)$. For the equivalent NC-CRC (in the second time-slot), set $X_{P2co}=X_{P2pr}=\phi$, and restrict the input distribution to correspond to independent rate-splitting and binning as in~\cite[(26)]{Rini09} to match the distribution corresponding to $\mathcal{R}_{D}$. Let the overall rate region thereby obtained be $\mathcal{R}^{in}_1$. Clearly, $\mathcal{R}^{in}_1 \subseteq \mathcal{R}$. Using the result for parallel Gaussian channels, it can be shown that for any choice of $\alpha$, the rate pair $(R_P, R_C)$ is achievable if
\begin{eqnarray}
&&\hspace{-.75cm} R_P \leq \frac{\alpha}{2} \log\left(1 + g_{PC}P_P\right), ~ (R^0_P, R^0_C) \in \mathcal{R}^{in}_{NC}, \label{eqn:R_in_1_bd1}\\
&&\hspace{-.75cm} R_C = \bar{\alpha} R^0_C, ~ R_{Pco} = \min \left\{ \frac{\alpha}{2} \log\left(1 + \frac{\eta_1 P_P}{1+\bar{\eta}_1P_P}\right), \right. \nonumber\\ 
&&\hspace{.75cm}\left. \frac{\alpha}{2} \log\left(1 + \frac{h_{PC}\eta_1 P_P}{1+h_{PC}\bar{\eta}_1P_P}\right)\right\} + \bar{\alpha} R^0_{Pco}, \label{eqn:R_in_1_bd2}\\
&&\hspace{-.75cm} R_{Ppr} \leq \frac{\alpha}{2} \log\left(1 + \bar{\eta_1} P_P\right) + \bar{\alpha}R^0_{Ppr}, \label{eqn:R_in_1_bd3}
\end{eqnarray}
\noindent where $\eta_1\in[0,1]$ is the power fraction allocated for transmitting $w_{Pco}$ in the first time-slot. Note that, given an $\eta$ value, $\eta_1$ may be chosen such that $\frac{\eta}{1+\bar{\eta}P_P} \leq \bar{\eta_1} \leq 1$. Then, comparing~\eqref{eqn:Protocol1_bd1}-\eqref{eqn:Protocol1_bd3} to~\eqref{eqn:R_in_1_bd1}-\eqref{eqn:R_in_1_bd3}  establishes that the region corresponding to Protocol 1 is contained in $\mathcal{R}^{in}_1$.

\begin{figure}[!ht]
\begin{center}
\includegraphics[scale=.45]{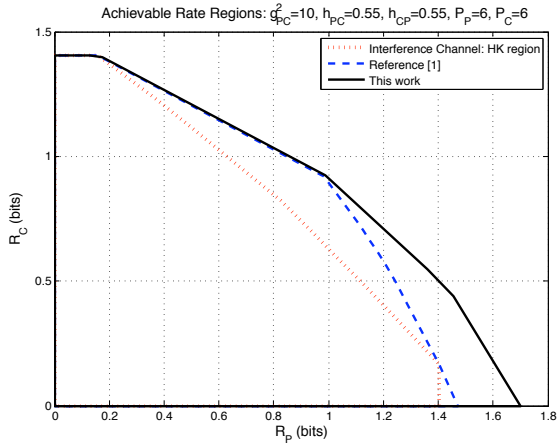}
\end{center}
\caption{Gaussian HD-CCRC: Weak interference for both cross-links.}
\label{fig:hd_weak_weak_interf}
\end{figure}

The inclusion of the rate region corresponding to Protocol 2 can be easily proved by considering the same coding structure and input distribution as used to obtain $\mathcal{R}^{in}_1$, with  one further restriction - the input distribution at $S_P$ for the first time-slot is given by $p(x_{P1co}|s=l)p(x_{P1pr}|s=l)$. This yields an achievable rate region $\mathcal{R}^{in}_2$ ($\subseteq\mathcal{R}$), that has exactly the same bounds as that for Protocol 2, except that the achievable rate region for the NC-CRC (during the second time-slot) is $\mathcal{R}^{in}_{NC} \supseteq \mathcal{R}_{DMT}$, thereby proving the above inclusion.

The rate region for Protocol 3 can be obtained by setting $S=t~\text{w.p.}~1, ~X_{P1co} = X_{P1pr} = T_{P1co} = T_{P1pr} = \phi$ in Theorem~\ref{thm:HD-causal_ach_rates}. Finally, the rate pair corresponding to Protocol 4 may be obtained by using a fixed listen-transmit schedule, and by setting $T_{P1co} = X_{P1co} = X_{P2co} = X_{P2pr} = U_{Cco} = U_{Cpr}=\phi$. As the four rate regions of~\cite{Devroye06} are contained in $\mathcal{R}$, the convex hull of these regions ($\mathcal{R}_0$) is also contained in $\mathcal{R}$ (cf. Remark~\ref{rem:convexity}). A numerical example comparing the Han-Kobayashi (HK) region, $\mathcal{R}_0$, and $\mathcal{R}$ is presented in Fig.~\ref{fig:hd_weak_weak_interf}. 


\end{document}